\documentstyle[12pt]{article}
\begin{document}
\baselineskip8mm
\title{\vspace{-3cm} Non-minimally coupled complex scalar field in
classical and quantum cosmology}
\author{A. Yu. Kamenshchik$^{1}$, \ I. M. Khalatnikov$^{2,3}$ and
\ A. V. Toporensky$^{4}$}
\date{}
\maketitle
\hspace{-6mm}$^{1}${\em Nuclear Safety Institute, Russian
Academy of Sciences, 52 Bolshaya Tulskaya, Moscow, 113191, Russia}\\
$^{2}${\em L. D. Landau Institute for Theoretical Physics,
Russian Academy of Sciences, Kosygin Street, Moscow, 117940,
Russia}\\
$^{3}${\em Tel Aviv University, Mortimer and Raymond Suckler
Institute of Advanced Studies,
 Ramat Aviv, 69978, Israel}\\
$^{4}${\em Sternberg
Astronomical Institute, Moscow University, Moscow, 119899, Russia}\\
\\
{\bf Abstract}

We investigate the cosmological model with complex scalar
self-interacting inflaton field non-minimally coupled to gravity.
The different geometries of the Euclidean classically forbidden
regions are represented. The instanton solutions of the corresponding
Euclidean equations of motion are found by numerical calculations.
These solutions give a rather non-trivial examples of real tunnelling
geometries. Possible interpretation of obtained results and  their
connection with inflationary cosmology is discussed.\\
\newpage

It is widely recognized that
inflationary cosmological models give a good basis for the
description of the observed structure of the Universe. Most of these
models include the so called inflaton scalar field  possessing
non-zero classical average value which provides the existence of
an effective cosmological constant on early stage of the
cosmological evolution.  In the series of recent papers [1--3] it was
suggested to consider the cosmological models with the complex scalar
field $\phi = x \exp (i \theta)$,
that is equivalent to the inclusion into the theory the new
quasi-fundamental constant -- charge of the Universe $Q$.

Appearance of this new constant essentially modifies the structure of
the Wheeler-DeWitt equation. Namely, the form of the superpotential
$U(x,a)$ displays now a new and interesting feature: the Euclidean
region, i.e. the classically forbidden region where $U > 0$, is
bounded by a closed curve in the minisuperspace $(x, a)$ for a large
range of parameters.
Thus, in contrast with the picture of ``tunneling from
nothing''[4] and with the ``no-boundary proposal'' for the wave
function of the Universe [5,6] we have Lorentzian region at the very
small values of cosmological radius $a$ and hence a wave can go into
Euclidean region from one side and outgo from the other.  These new
features of the model require reconsideration of traditional scheme
[4--6] and give some additional possibilities.

Here, it is necessary to stress that speaking about the ``Euclidean''
or ``classically forbidden'' regions one should understand that in
the case of the quantum gravity and cosmology these terms can be used
only in ``loose'' sense, because due to indefiniteness of the
supermetric the ``Euclidean'' region is not impenetrable for
Lorentzian trajectories. It is well-known that in the cosmological
models with inflaton scalar field the Lorentzian trajectories can
penetrate into Euclidean region just like as the trajectories
corresponding to the Euclidean equations of motion can leave
Euclidean region for the Lorentzian one (see, for example, Refs. [7--
14]. However, one can use such terms as Euclidean and Lorentzian
regions as people usually do investigating the tunneling type
processes in cosmology [7--14] and in instanton physics [15] even in
the case of indefinite supermetric. Moreover, one can give to the
term ``Euclidean region'' quite definite value as the region where
the points of the minimal contraction and maximal expansion of the
Universe can exist [16].

In recent years quite a few papers were devoted to the
investigation of the cosmological models with the non-minimal
coupling between inflaton scalar field and gravity [17--19]. On one
hand such models give a lot of opportunities for matching with the
observational data and ideas of particle physics, on the other hand
they can be treated as more consistent from the point of view of
quantum gravity [20].  This letter is devoted to the investigation of
the cosmological models with the complex scalar field non-minimally
coupled to gravity.

We shall consider the model with the following action:
\begin{eqnarray}
&& S = \int d^{4}x \sqrt{-g}\left(\frac{m_{Pl}^{2}}{16 \pi} (R - 2
\Lambda) + \frac{1}{2} g^{\mu \nu} \phi_{\mu}^{*} \phi_{\nu} \right.
\nonumber  \\
&&\left. + \frac{1}{2} \xi R \phi \phi^{*}- \frac{1}{2} m^{2} \phi
\phi^{*} - \frac{1}{4!} \lambda (\phi \phi^{*})^{2}\right).
\end{eqnarray}
Here $\xi$ is the parameter of non-minimal coupling  (we choose for
convenience the sign which is opposite to generally accepted),
$\lambda$ is the parameter of the self-interaction of the scalar
field, $\Lambda$ is cosmological constant, $m$ is the mass of the
scalar field. The complex scalar field $\phi$ can be represented in
the form
\[\phi = x \exp (i \theta)\]
We shall consider the minisuperspace model with the spatially
homogeneous variables $a$ (cosmological radius), $x$ and $\theta$.
Taking into account that the variable $\theta$ is cyclical one can
assume that the corresponding conjugate momentum is frozen and does
not subject to quantization [1--3]. In this case the only influence
of the variable $\theta$ on the dynamics of the system consists in
the appearance of the new constant of the model: classical charge
$Q$ which is defined as
\[Q = a^{3} x^{2} \dot{\theta}.\]
Let us notice that the another approach to the complex scalar
field in cosmology was discussed in [21] mainly in the context
of wormhole solutions.

The super-Hamiltonian constraint equation in minisuperspace
variables $(a,x)$ has the following form:
\begin{eqnarray}
&&-\frac{p_{a}^{2}}{(2\pi^{2})^{2} 24 a
\left(\frac{m_{Pl}^{2}}{16\pi} +\frac{\xi x^{2}}{2} + 3 \xi^{2}
x^{2}\right)} -\frac{p_{x} p_{a} \xi x}{(2\pi^{2})^{2} 2 a^{2}
\left(\frac{m_{Pl}^{2}}{16\pi} +\frac{\xi x^{2}}{2} + 3 \xi^{2}
x^{2}\right)}\nonumber \\
&&\frac{p_{x}^{2}}{(2\pi^{2})^{2} 2 a^{3}}
\frac{\left(\frac{m_{Pl}^{2}}{16\pi} +\frac{\xi x^{2}}{2} + \right)}
{\left(\frac{m_{Pl}^{2}}{16\pi}
+\frac{\xi x^{2}}{2} + 3 \xi^{2} x^{2}\right)}
-U(a,x) = 0,
\end{eqnarray}
where $p_{a}$ and $p_{x}$ are the momenta canonically conjugate to
$a$ and $x$ respectively and superpotential $U(a,x)$
\footnote {In papers [1,2] instead of superpotential the
effective mass $m^{2}(\ln a, x) = - U(a, x)/a $ was used
and the condition of ``euclidity''
was :  $m^{2}(\ln a, x) < 0$}
looks as follows:
\pagebreak
\begin{eqnarray} &&U(a,x) = a \left(\frac{m_{Pl}^{2}}{16 \pi}(6 - 2
\Lambda a^{2})) + 3 \xi x^{2}\right. \nonumber \\
&&\left.
-\frac{Q^{2}}{a^{4} x^{2}} - \frac{1}{2} m^{2} x^{2} a^{2} -
\frac{1}{24} \lambda x^{4} a^{2}\right).  \end{eqnarray}

Resolving equation $U = 0$ we can get the form of the Euclidean region
in  the plane of minisuperspace variables $(a,x)$. It is interesting
to compare the form of these regions for different values of
parameters in the action (1). All cases are represented on Fig. 1.
In the simplest case then $Q = \Lambda = \lambda = \xi = 0$ we have
non-compact Euclidean region bounded by hyperbolic curve $x = \sqrt
{\frac{3}{4 \pi}} \frac{m_{Pl}}{m a}$ (see Fig. 1a). Inclusion of
the cosmological term $\Lambda \neq 0$ implies the closing of the
Euclidean region ``on the right'' at $a = \sqrt{\frac{3}{\Lambda}}$
(see Fig. 1b).  Inclusion of the non-zero classical charge of the
scalar field $Q \neq 0$ implies the closing of the Euclidean region
``on the left'' and we have obtained ``banana-like'' structure of
this region [1--3] (see Fig. 1c). After the inclusion of the small
term describing the non-minimal coupling between scalar field and
gravity $(\xi \neq 0)$ we obtain the second Euclidean region in the
upper left corner of our picture (see Fig. 1d). This new region is
non-compact and unrestricted from above.
While increasing the value of the parameter $\xi$ this
second Euclidean region drops down and at some value of $\xi$
(it is easy to find this value of $\xi$:
$\xi = \frac{16 \pi^{2} m^{4} Q^{2}}{27 m_{Pl}^{4}}$)
joins with the first banana-like Euclidean region. The boundary
of this unified region is partially convex, partially concave (see
Fig. 1e) and after further increasing of $\xi$ it becomes convex (see
Fig. 1f).  After  inclusion of self-interaction of the scalar field
$\lambda \neq 0$ we can have, depending on the values of the
parameters  $Q, \lambda, \xi$ and $m$, various geometrical
configurations of the Euclidean regions. It is easy to estimate the
condition of closing of the Euclidean region from above is $\xi <
\left(\frac{Q \lambda}{48}\right)^{2/3}$. Now  remembering that the
condition of the existence of two non-connected Euclidean regions is
$\xi < \frac{16 \pi^{2} m^{4} Q^{2}}{27 m_{Pl}^{4}}$ one can see that
we have three options. First, if
$\xi < \frac{16 \pi^{2} m^{4} Q^{2}}{27 m_{Pl}^{4}}$
we have only one closed banana-like Euclidean region (see Fig. 1g).
Second, if
\[\frac{16 \pi^{2} m^{4} Q^{2}}{27 m_{Pl}^{4}} >
\left(\frac{Q \lambda}{48}\right)^{2/3}\]
and
\[\frac{16 \pi^{2} m^{4} Q^{2}}{27 m_{Pl}^{4}} > \xi >
\left(\frac{Q \lambda}{48}\right)^{2/3},\]
we shall have two non-connected Euclidean regions : banana-like one
and ``bag-like'' Euclidean region with an infinitely long narrow
throat (it looks again as in Fig. 1d, but the curves bounding the
upper Euclidean region are asymptotically clinging to the ordinate
axis).

Third, if $\xi > \left(\frac{Q \lambda}{48}\right)^{2/3}$
and $\xi > \frac{16 \pi^{2} m^{4} Q^{2}}{27 m_{Pl}^{4}}$ we shall
have one open above bag-like Euclidean region which again has an
infinitely long narrow throat (see Fig 1h). Thus, we have seen that
inclusion of the charge $Q$, non-minimal coupling $\xi \neq 0$ and
self-interaction of the scalar field implies a large variety of
possible geometries of Euclidean regions in minisuperspace.

Let us now write down the equations of motion for our minisuperspace
system including gravity and scalar field. Varying the action (1)
in respect to $a$ and $x$ and choosing for the lapse function
the value $N = 1$ we obtain
\begin{eqnarray}
&&\frac{m_{Pl}^{2}}{16 \pi}\left(\ddot{a} + \frac{\dot{a}^{2}}{2 a}
+ \frac{1}{2 a} - \frac{\Lambda a}{2}\right)
+\frac{\xi \dot{a}^{2} x^{2}}{4 a} + \frac{\xi \ddot{a} x^{2}}{2}
+ \xi x \dot{x} \dot{a} + \frac{\xi \dot{x}^{2} a}{2}\nonumber \\
&&+\frac{\xi x \ddot{x} a}{2} + \frac{\xi x^{2}}{4 a}
+\frac{a \dot{x}^{2}}{8}
-\frac{m^{2} x^{2} a}{8} + \frac{Q^{2}}{4 a^{5} x^{2}} -
\frac{\lambda x^{4} a}{96} = 0
\end{eqnarray}
and
\begin{eqnarray}
&&\ddot{x} + \frac{3 \dot{x} \dot{a}}{a} - \frac{6 \xi x \ddot{a}}{a}
- \frac{6 \xi \dot{a}^{2} x}{a^{2}}\nonumber \\
&&-\frac{6 \xi x}{a^{2}} + m^{2} x - \frac{2 Q^{2}}{a^{6} x^{3}}
+\frac{\lambda x^{3}}{6} = 0.
\end{eqnarray}
Besides we can write down the first integral of motion of our
dynamical system which can be obtained from the super-Hamiltonian
constraint (2):
\begin{eqnarray}
&&-\frac{3}{8 \pi} m_{Pl}^{2} a \dot{a}^{2} - 3 \xi a \dot{a}^{2}
x^{2} -6 \xi x \dot{x} \dot{a} a^{2}\nonumber \\
&&+\frac{a^{3}}{2} \dot{x}^{2} - U(a,x) = 0.
\end{eqnarray}

It is obvious that Euclidean counterparts of the
Eqs.  (4)--(6) can be obtained by the changing of sign before terms
containing time derivatives.

Numerically integrating the Euclidean analog of the system of
equations (4)--(6) we can investigate the question about the presence
of instantons between the solutions of these equations in Euclidean
region. Under instantons we shall understand solutions of Euclidean
equations of motion which have vanishing velocities on the boundaries
of Euclidean region (in the case when the boundary of the Euclidean
region is partially convex and partially concave the condition of the
extremum of action does not require with necessity the vanishing of
velocities on the boundary between regions and the so called
non-trivial instantons can exist [1,2], but we do not study this
opportunity in this letter).

It appears that in the case when we have two non-connected
Euclidean regions, a closed banana-like one and an open region in the
upper left corner of the picture which arises due to the presence of
the non-minimal coupling constant $\xi$, then the situation in the
closed region is qualitatively the same as in the case with
minimal coupling which was considered in [1,2]. Namely, inside the
banana-like region we have an instanton solution which begins in one
point of the
boundary of this region with zero initial velocities $\dot{a}$ and
$\dot{x}$ and ends at another point of the boundary with zero final
velocities. Naturally, this solution corresponds to the local maximum
value of action of our dynamical system.
As far as concerned an open region
we observe that an action is
monotonically increasing while the initial value of $x$ is increasing
and there are not instantons at all.

In the case when closed and open Euclidean region are glued together
and we have only one open Euclidean region there are two options :
we can have two instantons or no instanton depending
on parameters (see Fig. 2).  In the first case we can see that
one of these instantons (in the
bottom part of the open Euclidean region) corresponds to local
maximum of action while the second instanton corresponds to local
minimum of action (see Fig. 2a). At the same time taking the growing
initial values of $x$ on the boundary of the Euclidean region under
consideration (above the second instanton) we shall get the Euclidean
trajectories with the growing action. In the second case we have only
trajectories with the monotonically increasing action and instantons
do not exist (see Fig. 2b).

Now we are in a position to say some words about the interpretation
of the obtained results. All the models describing ``the birth of the
Universe from nothing'' in the form of tunnelling wave function [4]
or due to no-boundary mechanism [5,6] are grounded on the assumption
that it is  possible to provide the analytical continuation from
Lorentzian geometry to Euclidean one [11--16]. Here, it is necessary
to stress that there is essential difference between tunnelling
calculations in non-relativistic quantum mechanics and those in
quantum gravity and cosmology. In non-relativistic quantum mechanics
the metric describing the structure of the kinetic term in Lagrangian
is positive-definite and there are well-characterized conditions
under which the solution connecting two points is real or complex.
Moreover, the boundary between Euclidean and Lorentzian regions is
sharply determined by the condition $U = 0$. At these boundary
conditions all the velocities should be equal to zero. With the
indefinite supermetric arising in quantum gravity and cosmology
things are more complicated. Really, in this case the
super-Hamiltonian constraint according to which kinetic term should
be equal to superpotential does not imply the simultaneous vanishing
of all the velocities at the naive boundary of the Euclidean and
Lagrangian region which is defined by the condition $U = 0$.
Moreover, such solutions which have vanishing velocities on the
boundary do not exist in some models. It is enough to mention the
comparatively simple Hawking's model with real minimally-coupled
scalar field without interaction [6], where it is impossible to have
solutions possessing the simple matching between Euclidean and
Lorentzian regions [6]. The models in which the one can match the
Euclidean and Lorentzian trajectories at some points where all the
velocities turn to zero (so called real
tunnelling geometries) are usually the different theories without
matter but with the cosmological constant and they investigated in
some detail [11]. In the more realistic cases we encounter the
necessity to consider complexified classical trajectories in the
minisuperspace [9--14] which makes things much more complicated.
The very notion of the boundary between Euclidian and Lorentzian
becomes fuzzy. The complete theory of such complexified transition
between Euclidean and Lorentzian geometries has not yet elaborated
and these questions certainly deserve further investigation. However,
the simple instantons which were found in papers [1,2] for the closed
Euclidean regions and those found in the present paper for open
Euclidean regions are of some interest because they supply us with
the examples of non-trivial real tunnelling geometries in the models
with matter which interact with gravity in a very non-trivial way. In
 spite of the question about the interrelation between real and
complex tunnelling geometries are not clear as was explained above,
the former has at least one advantage - a rather simple and apparent
physical interpretation which could be reduced to different version
of birth of the Universe from nothing and is analogous to usual
tunneling transition in quantum mechanics. Taking this fact into
account it is interesting to continue the instantons found in our
model into Lorentzian region and look at their behaviour. It is
remarkable that at least some of them have quasi-inflationary
behaviour (see Fig. 3).

In the conclusion let us note some remarkable features of the
non-minimally coupled model considered in this letter. First, due to
the opportunity to have an open Euclidean region at large values of
scalar field $x$ (see Fig. 1e,f,g,i,j) we can exclude classical
Lorentzian trajectories which can go around closed Euclidean region
as it was in the minimally coupled complex scalar field model [1,2].
Such trajectories have a comparable small probabilities but due to
their abundance the contribution of the trajectory originating from
instanton could be suppressed.

Second, the  instantons which were
found in the closed banana-like region in the minimally coupled model
have their end points near the end of ``banana'' whose localization
is determined by the value of the cosmological constant $\Lambda$.
Thus, the initial parameters of inflation are up to some extent
determined by the cosmological constant $\Lambda$ while according to
the very spirit of the inflationary cosmology they should be
determined by the inflaton scalar field. Here, in the cases when we
one open Euclidean region (see Fig. 1f,j)
we can find the Lorentzian Universes which are born from the
instanton which corresponds to the minimum of action. The parameters
of this Lorentzian Universe
are practically independent of $\Lambda$ and are determined by the
values of $Q,\lambda,m$ and $\xi$, i.e.  the numbers, describing the
properties of inflaton scalar field in the full correspondence with
our expectations. It is not difficult to find such values of
parameters at which this minimal value of action is close to maximal
one and the relative probability of birth of such an Universe is
rather high.

{\bf Acknowledgement}
The research described in this publication was also made possible in
part by grant No MAE000 from the International Science Foundation.
A.Yu.K.  is also partially supported by Russian Foundation for
Fundamental Researches through grant No 94-02-03850-a and by Russian
Research Project ``Cosmomicrophysics''.
\\ \\
\begin{description}
\item[{[\rm 1]}] I.M. Khalatnikov and A.
Mezhlumian, Phys. Lett. A 169 (1992) 308.
\item[{[\rm 2]}] I.
M. Khalatnikov and P. Schiller, Phys. Lett. B 302 (1993) 176.
\item[{[\rm 3]}]
L. Amendola, I.M. Khalatnikov, M. Litterio and F. Occhionero,
Phys. Rev. D 49 (1994) 1881.
\item[{[\rm 4]}]
A. Vilenkin, Phys. Lett. 117B (1982) 25; Phys. Rev. D
27 (1983) 2848; Phys. Rev. D 30 (1984) 509;
Phys. Rev. D 37 (1988) 888; A.D. Linde,
Zh. Eksp. Teor. Fiz. 87 (1984) 369  [Sov. Phys. JETP 60
 (1984) 211]; Ya.B. Zeldovich and A.A. Starobinsky, Pis'ma Astron.
Zh. 10 (1984) 323  [Sov. Astron. Lett. 10 (1984) 135];
V. A. Rubakov, Phys. Lett. 148 B (1984) 280.
\item[{[\rm 5]}]
J.B. Hartle and S.W. Hawking, Phys. Rev. D 28 (1983) 2960.
\item[{[\rm 6]}]
S.W. Hawking, Nucl. Phys. B 239 (1984) 257.
\item[{[\rm 7]}]
D.N. Page, Class. Quantum Grav. 1 (1984) 417.
\item[{[\rm 8]}]
V.A. Belinsky, L.P. Grishchuk, Ya.B. Zel'dovich and
I.M. Khalatnikov, J. Exp. Theor. Phys. 89 (1985) 346;
V.A. Belinsky and I.M. Khalatnikov, J. Exp. Theor. Phys. 93 (1987)
784.
\item[{[\rm 9]}]
J.J. Halliwell and J.B. Hartle, Phys. Rev. D 41 (1990) 1815.
\item[{[\rm 10]}]
J.J. Halliwell and J. Louko, Phys. Rev. D 39 (1989) 2206;
40 (1990) 1868; 42 (1990) 3997.
\item[{[\rm 11]}]
G.W. Gibbons and J.B. Hartle, Phys. Rev. D 42 (1990) 2458.
\item[{[\rm 12]}]
G.W. Gibbons and H.-J. Pohle, Nucl. Phys. B 410 (1993) 117.
\item[{[\rm 13]}]
A.O. Barvinsky, Phys. Reports 230 (1993) 237.
\item[{[\rm 14]}]
A.O. Barvinsky and A.Yu. Kamenshchik, Phys. Rev. D 50 (1994) 5093.
\item[{[\rm 15]}]
G.V. Lavrelashvili, V.A. Rubakov, M.S. Serebryakov and P.G. Tinyakov,
Nucl. Phys. B 342 (1990) 98;
V.A. Rubakov and P.G. Tinyakov, Nucl. Phys. B 342 (1990) 430.
\item[{[\rm 16]}]
A.Yu. Kamenshchik, I.M. Khalatnikov and A.V. Toporensky, work in
progress.
\item[{[\rm 17]}]
D.S. Salopek, J.R. Bond and J.M. Bardeen, Phys.
Rev. D 40 (1989) 1753.
\item[{[\rm 18]}]
R. Fakir and W.G. Unruh, Phys. Rev. D 41 (1990) 1783;
R. Fakir, S. Habib and W.G. Unruh, Ap. J. 394 (1992) 396;
R. Fakir and S. Habib, Mod. Phys. Lett. A 8 (1993) 2827.
\item[{[\rm 19]}]
A.O. Barvinsky and A.Yu. Kamenshchik, Class. and Quantum Grav.
7 (1989) L181; A.Yu. Kamenshchik, Phys. Lett. B 316 (1993) 45;
A.O. Barvinsky and A.Yu. Kamenshchik, Phys. Lett. B
232 (1994) 270.
\item[{[\rm 20]}]
A.O. Barvinsky, A.Yu. Kamenshchik and I.P. Karmazin, Phys. Rev. D
 48 (1993) 3677.
\item[{[\rm 21]}]
K. Lee, Phys. Rev. Lett. 61 (1988) 263;
J.D. Brown, C.P.Burgess, A Kshisagar, B.F. Whiting and J.W. York,
Nucl. Phys. B 326 (1989) 213;
S. Coleman, Nucl. Phys. B 329 (1990) 387;
J. Twamley and D.N. Page, Nucl. Phys. B 378 (1992) 247;
K. Lee, Phys. Rev. D 50 (1994) 5333.
\end{description}
\newpage
{\bf Captions to Figures}.
FIG. 1.  The boundaries of Euclidean regions are represented on this
figure. The absolute value of scalar inflaton field $x$ is measured
in Planck masses, while cosmological radius $a$ in Planck lengths.
The values of parameters are the following: (a) $Q=\Lambda=\lambda=
\xi=0, m^{2}=m_{Pl}^{2}/32\pi$; (b) $Q=\lambda=
\xi=0, m^{2}=m_{Pl}^{2}/32\pi, \Lambda=0.2m_{Pl}^{2}/16\pi$;
(c) $\lambda=
\xi=0, m^{2}=m_{Pl}^{2}/32\pi, \Lambda=0.5m_{Pl}^{2}/16\pi,
Q=15.5$;
(d) $\lambda=0, m^{2}=m_{Pl}^{2}/32\pi, \Lambda=0.5m_{Pl}^{2}/16\pi,
Q=15.5, \xi=0.15$;
(e) $\lambda=0, m^{2}=m_{Pl}^{2}/32\pi, \Lambda=0.5m_{Pl}^{2}/16\pi,
Q=15.5, \xi=1.15$;
(f) $\lambda=0, m^{2}=m_{Pl}^{2}/32\pi, \Lambda=0.5m_{Pl}^{2}/16\pi,
Q=15.5, \xi=8.15$;
(g) $\lambda=0.175, m^{2}=m_{Pl}^{2}/32\pi,
\Lambda=0.34m_{Pl}^{2}/16\pi, Q=15.5, \xi=1.175$;
(h) $\lambda=0.1, m^{2}=m_{Pl}^{2}/32\pi,
\Lambda=0.075m_{Pl}^{2}/16\pi, Q=15.5, \xi=50.1$
\\
FIG. 2. (a) Classical Euclidean trajectories including instantons;
(b) Classical Euclidean trajectories in the case when instantons are
absent.
\\
FIG. 3. Classical Lorentzian trajectories matched to instantons on
the boundary of Euclidean region.
\end{document}